\documentclass[reprint,twocolumn,superscriptaddress,nofootinbib,aps,pre,floatfix]{revtex4-1}
\pdfoutput=1
\usepackage{graphicx}
\usepackage[english]{babel}
\usepackage{amsmath}
\usepackage{amssymb}
\usepackage[usenames,dvipsnames]{color}
\usepackage{url}
\usepackage{bbm}
\usepackage{multirow}
\usepackage{hyperref}

\usepackage{bm}
\usepackage[caption=false]{subfig}
\usepackage{float}
\usepackage{multirow}
\usepackage{natbib}
\usepackage{mdwlist}
\usepackage{threeparttable}
\usepackage{booktabs}
\usepackage{tabularx}

\usepackage[top=4cm, bottom=3.8cm, left=2.5cm, right=2.5cm]{geometry}
\usepackage[title]{appendix}

\usepackage{amsmath,amssymb,amsfonts,upref,endnotes}
\usepackage{amsthm}
\usepackage{graphicx}
\usepackage{color}

\newcounter{listnumber}
\newcommand{\Ron}[0]{\mathcal{R}_{\text{ON}}}
\newcommand{\Roff}[0]{\mathcal{R}_{\text{OFF}}}
\newcommand{\zMax}[0]{z_{\text{max}}}

\begin{document}
\title{Window functions and sigmoidal behaviour of memristive systems}
\author{Panayiotis S. Georgiou} \email[]{ps.georgiou@imperial.ac.uk}
\affiliation{Department of Bioengineering, Imperial College London, 
    London SW7 2AZ, United Kingdom}
\author{Sophia N. Yaliraki} 
\affiliation{Department of Chemistry, Imperial College London, 
    London SW7 2AZ, United Kingdom}
\author{Emmanuel M. Drakakis}
\affiliation{Department of Bioengineering, Imperial College London, 
    London SW7 2AZ, United Kingdom}
\author{Mauricio Barahona}\email[]{m.barahona@imperial.ac.uk}
\affiliation{Department of Mathematics, Imperial College London,  
    London SW7 2AZ, United Kingdom}
    

\begin{abstract}
	A common approach to model memristive systems is to include empirical window functions to describe edge effects and non-linearities in the change of the memristance. We demonstrate that under quite general conditions, each window function can be associated with a sigmoidal curve relating the normalised time-dependent memristance to the time integral of the input. Conversely, this explicit relation allows us to derive window functions suitable for the mesoscopic modelling of memristive systems from a variety of well-known sigmoidals. Such sigmoidal curves are defined in terms of measured variables and can thus be extracted from input and output signals of a device and then transformed to its corresponding window. We also introduce a new generalised window function that allows the flexible modelling of asymmetric edge effects in a simple manner. 
\end{abstract}

\maketitle


\section{Introduction}

The memristor, which was originally defined by Leon Chua in 1971 based on symmetry arguments, is the one-port element that relates charge and flux-linkage~\cite{Chua1971}. The memristor completes the tetrad of ideal passive elements together with the resistor, capacitor and inductor. For decades, memristor research was mostly of theoretical interest until in 2008 researchers at Hewlett Packard (HP) fabricated a nano-scale device whose behaviour was described with a memristor model~\cite{Strukov2008}. The experimental and theoretical interest in memristors has since grown rapidly, motivated by their possible use to improve existing circuits, as well as enabling novel applications. Their intrinsic non-volatile memory~\cite{Jo2009}, in combination with their low-power consumption~\cite{Schindler2008}, nano-scale size~\cite{Waser2009}, high switching speed~\cite{Torrezan2011} and synapse-like behaviour~\cite{Jo2010b},  are all desirable properties for potential uses in computer memories, as well as reconfigurable, neuromorphic and learning/adaptive circuits~\cite{Strukov2010}.

A wide range of resistance-switching devices have been classified as memristive. In most cases, the device exhibits two limiting resistance states (`high' and `low') with an electrically-induced change in conductance (i.e., by applying voltage or current). However, the underlying mechanisms for these induced changes are complex and differ significantly among devices~\cite{Waser2009}. As a result, a myriad of models constructed from microscopic physical mechanisms or based on fitting to macroscopic experimental measurements (or combinations of both) have been proposed~\cite{Yang2008,Strukov2009a,Strukov2009b,Yu2010b,Pickett2009,Kvatinsky2013,Vourkas2015,Ascoli2015}. As an alternative to such detailed models, macroscopic empirical models have been developed to capture input-output responses, and in particular, properties related to hysteresis, memory, edge effects, and nonlinearities in the change of the memristance. In this setting, empirical \textit{window functions} are commonly used to encapsulate the nonlinear mechanisms that affect the change in the memristance and limit its range of values. Following the window function introduced by the HP group~\cite{Strukov2008}, several groups have since introduced window functions capturing different observed properties~\cite{Yogesh2009,Biolek2009,Benderli2009,Prodromakis2011,Corinto2012,Takahashi2014}. For a review of existing window functions and related models the reader may refer to Refs.~\cite{Eshraghian2012,Ascoli2013,Linn2014}.

Here we study the mathematical properties of window functions in terms of the associated saturating behaviour of the memristor. In particular, we show that, under some general assumptions, a window function leads to a sigmoidal in the observable signal of the system. This sigmoidal function can thus be obtained from input-output measurements and used to infer the underlying window function from the data. We also introduce a generalised window function with adjustable parameters that allow us to flexibly describe asymmetric boundary effects in the device, and we use it to illustrate the mathematical link between window and sigmoidal functions in simulated data. Finally, we exploit the extensive results on sigmoidal functions to construct a list of possible window functions associated with well-known sigmoidals, which can be used to model input-output memristive responses.

\begin{figure}[!tb]
\begin{center}
\includegraphics[]{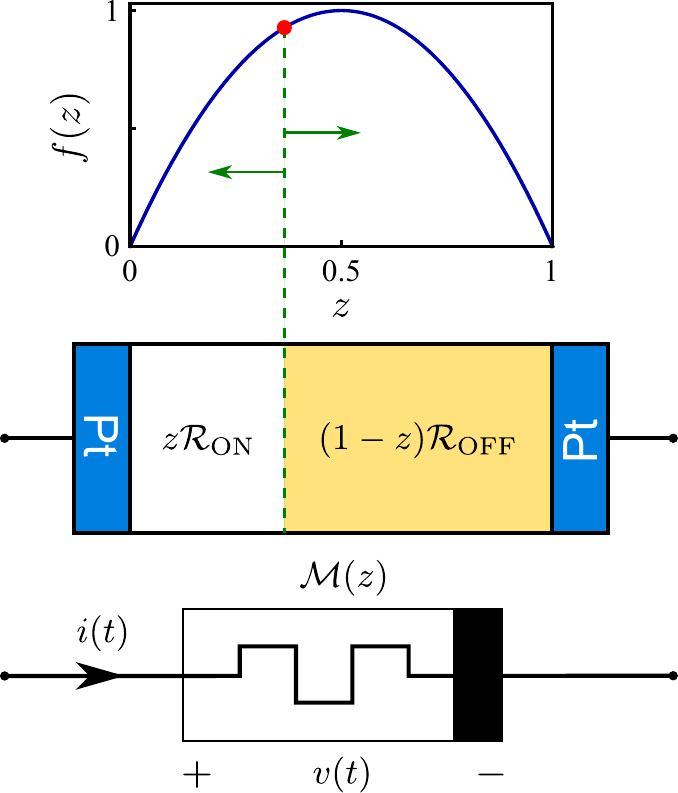}
\caption[]
{
	Memristive model and window function. When an input (current or voltage) is applied, the boundary between the doped and undoped regions moves, inducing a change in the memristance, i.e., a change in $z$. The edge effects are represented by the function $f(z)$:  as $z$ approaches either of its limiting values $z=0$ or $z=1$, it becomes increasingly harder for the memristance to change. The device saturates at $z=0$ or $z=1$ corresponding to the limiting resistances $\Roff$ or $\Ron$, respectively. The device schematic is adapted from Strukov \textit{et al}~\cite{Strukov2008}.
}
\label{fig:memristor_window_illustration}
\end{center}
\end{figure}

\section{Memristors and Memristive Systems}
\label{sec:Memristors and Memristive Systems}

A memristor is defined by the constitutive relation:~\cite{Chua1971}
\begin{equation}
\label{eq:constitutive}
f_{\mathcal{M}}(q,\varphi)=0,
\end{equation}
which relates the charge $q$ and flux-linkage $\varphi$. The memristor is said to be charge-controlled if it is possible to express the implicit relation~\eqref{eq:constitutive} as an explicit function of the charge: $\varphi=\hat{\varphi}(q)$. It then follows that 
\begin{equation}
 \label{eqn:memristor_definition}
 v=\mathcal{M}(q) \, i(t),
\end{equation}
which describes the output $v$ of the charge-controlled memristor with memristance $\mathcal{M}(q)$ under the time-varying current input
\begin{equation}
 \label{eqn:current=dq_dt}
 i = \dfrac{dq(t)}{dt}.
\end{equation} Note that the memristance $\mathcal{M}(q)=d\varphi/dq$ is a function only of the time integral of the input and is equal to the ratio between voltage and current (i.e., output and input) at each time, $t$. The description of a flux-controlled memristor, defined by the explicit function $q=\hat{q}(\varphi)$, follows analogous arguments. 

An \textit{ideal memristor}, as defined originally by Chua~\cite{Chua1971}, is characterised by a unique and time-invariant $q-\varphi$ function with the following properties~\cite{Georgiou2014,Georgiou2013}:
\begin{enumerate*}
\stepcounter{listnumber}
\renewcommand{\theenumi}{\Alph{listnumber}.\arabic{enumi}}
\renewcommand{\labelenumi}{\Alph{listnumber}.\arabic{enumi}}
 \item \label{list:ideal-memristor-properties2} nonlinear    \label{A.1}
 \item \label{list:ideal-memristor-properties3} continuously differentiable  \label{A.2} 
 \item \label{list:ideal-memristor-properties4} strictly monotonically increasing,  \label{A.3}
\end{enumerate*}
and on the $i-v$ plane its canonical form is obtained from (\ref{eqn:memristor_definition}) and (\ref{eqn:current=dq_dt}). Under these conditions, the charge-controlled $\hat{q}(\varphi)$ and flux-controlled $\hat{\varphi}(q)$ functions are invertible, and these two representations of the memristor are equivalent~\cite{Georgiou2012,Georgiou2012a,Drakakis2010}. 

To enable the modelling of a broader class of systems with memristor-like properties, Chua and Kang later introduced  the concept of \textit{memristive systems}~\cite{Chua1976}, in which the memristance may depend not only on internal state variables of the system, but also can be a direct function of the input drive and of time.

In this work, we consider a particular form of memristive systems that has been used widely to model recent experiments~\cite{Eshraghian2012}. (Henceforth we focus on the current-driven system, but our work applies equally to the voltage-driven case). The model assumes that the memristance corresponds to a mixed resistor formed by a low resistance $\Ron$ and a high resistance $\Roff$ ($\Roff > \Ron$) in series, with the total effective resistance equal to the weighted sum of the two:
\begin{equation}
 \label{eqn:effective_resistance}
 \mathcal{M}(z) = z \Ron  +(1-z) \Roff,
\end{equation}
where $z \in [0,1]$ is an internal (dimensionless) state variable whose dynamics represents the nonlinear change in the memristance. The equations for the memristive system are then:
\begin{subequations}
\label{eqn:memristive_system_definition_special}
  \begin{align}
    \dot{z} &= \alpha f(z)\,i(t)    \label{eqn:memristive_system_definition_special_state} \\
          v &= \mathcal{M}(z)\,i(t) \label{eqn:memristive_system_definition_special_output},      
  \end{align}
\end{subequations}
where $\alpha$ is a constant of appropriate dimensions (here Coulomb$^{-1}$), and $f(z)$ is the dimensionless \textit{window function} that encapsulates boundary effects and non-linearities in the change of the memristance. Such non-linearities may arise because of different underlying mechanisms, for example, due to nonlinear ionic drift. In the original HP memristor model~\cite{Strukov2008}, the internal state variable $z$ was associated with the position of the front separating the doped and undoped regions in the TiO$_2$ nanoscale structure, and a window $f(z)$ with parabolic shape was adopted to represent edge effects (Figure~\ref{fig:memristor_window_illustration}). 

To avoid notational confusion, we remark that in his recent new classification of memristors~\cite{Chua2014,Chua2015}, Leon Chua defines the devices characterised by~\eqref{eqn:memristive_system_definition_special} as \textit{ideal generic memristors}. Additionally, in those papers, Chua relaxes the requirements for ideal memristors allowing them to be not only passive devices (as originally defined in~\cite{Chua1971}) but also active devices. In this work, however, we follow strictly the definitions given by Chua in the original Refs.~\cite{Chua1971,Chua1976} and restrict our analysis to (passive) ideal memristors, since~(\ref{A.1})--(\ref{A.3}) constitute the minimal set of properties giving rise to memristive behaviour (e.g. hysteresis, memory, zero-crossing)~\cite{Georgiou2013}. Under the new classification, the system in (\ref{eqn:memristive_system_definition_special}) falls into the class of ideal generic memristors. It was recently shown~\cite{Chua2014,Chua2015}, as demonstrated here, that under certain conditions this class reduces to ideal memristors. For the sake of clarity, we will henceforth follow strictly the original terminology in~\cite{Chua1971,Chua1976}. We remark that our aim here is not to introduce a new class of memristors, but to explore the full implications of considering fundamental properties of strictly passive ideal memristors (see also~\cite{Riaza2011,Jansen2015}). 

\section{From Window Functions to Sigmoidals}
\label{sec:from window functions to sigmoidals}

\subsection{Properties of window functions and a new generalised window}
\label{sec:A new window function}

The window function $f(z)$ in~\eqref{eqn:memristive_system_definition_special_state} is introduced as a means to describe a variety of effects, including edge effects and nonlinearities in the drift, and is either obtained from microscopic considerations~\cite{Strukov2009b,Pickett2009,Yu2010b,Kvatinsky2013} or, more empirically, from macroscopic arguments~\cite{Strukov2008,Benderli2009,Biolek2009,Yogesh2009,Prodromakis2011,Corinto2012,Takahashi2014}. In its very simplest form, $f(z)=1$, as in the first HP memristor model, which assumed linear drift and neglected boundary effects~\cite{Strukov2008}. However, already in Ref.~\cite{Strukov2008}, a second model with a parabolic window was proposed to reflect boundary effects and drift nonlinearities. Subsequently, a variety of window functions have been introduced based on different semi-empirical principles to reflect different underlying properties~\cite{Yogesh2009,Biolek2009,Benderli2009,Prodromakis2011,Corinto2012,Takahashi2014}.

Most of the proposed window functions~\cite{Strukov2008,Yogesh2009,Benderli2009,Prodromakis2011} (although not all~\cite{Biolek2009,Corinto2012,Takahashi2014}) share a common mathematical description given by the following specifications:
\begin{enumerate*}
 \stepcounter{listnumber}
 \renewcommand{\theenumi}{\Alph{listnumber}.\arabic{enumi}}
 \renewcommand{\labelenumi}{\Alph{listnumber}.\arabic{enumi}}
  \item \label{list:window-prop1} $f(z):[0,1] \rightarrow [0,1]$
  \item \label{list:window-prop2} $f(0)=f(1)=0$
  \item \label{list:window-prop3} single maximum $f(z_{\text{max}})=1$
  \item \label{list:window-prop4} continuous differentiability of $f(z)$,
\end{enumerate*}
i.e., the window is a strictly concave function in the normalised internal state variable $z\in[0,1]$ with two roots at $z=0$ and $z=1$  and a single maximum at $z=z_\text{max}$. 

Within this general form, we combine and extend several features of previously proposed window functions into the following generalised window:
\begin{subequations}
\label{eqn:window-new}
\begin{align}
f(z;r,p)&=\left[ \psi(r) \, z \, (1-z^{r-1})\right]^p, \quad r \neq 1 \label{eqn:window-new-function} \\
\psi(r)&=\frac{1}{\zMax (1-\zMax^{r-1})} \label{eqn:window-new-psi}
\end{align}
\end{subequations}
where the parameters $p\in(0,1]$ and $r\in\mathbb{R}^+-\left\lbrace 1\right\rbrace$ offer the flexibility to adjust the asymmetry and flatness of the window. Here $\psi(r)$ is a normalization constant such that $f(\zMax;r,p)=1$, and the maximum is at $\zMax=r^{1/(1-r)}$. Note that \eqref{eqn:window-new} satisfies the properties \eqref{list:window-prop1}-\eqref{list:window-prop4}.

\begin{figure*}[!tb]
\begin{center}
\includegraphics[width=\textwidth]{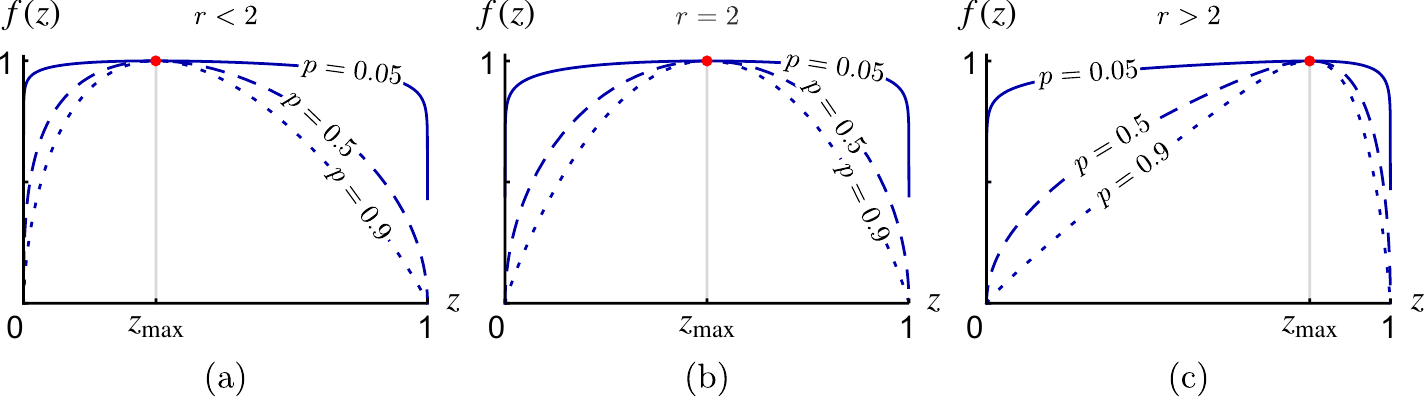}
\caption[]
{ 
	The effect of the parameters $r$ and $p$ on the window~\eqref{eqn:window-new}. The parameter $r$ controls the position of the maximum $\zMax$ and, as a result, the asymmetry of the window: (a) $0<r<2$, left skewed; (b) $r=2$, centred; (c) $r>2$, right skewed. The parameter $p$ controls the flatness of the region around $\zMax$, hence the depth of the boundary effects.  This is illustrated in all three plots where the window is plotted for different $p$ values for each value of $r$.  As  $p$ approaches $0$ the window becomes increasingly flatter. The specific values used are $r=0.8$ in (a) and $r=12.2$ in (c).
}
\label{fig:window_m_effect}
\end{center}
\end{figure*}

The window function~\eqref{eqn:window-new} provides a generalization of previous windows~\cite{Yogesh2009,Prodromakis2011,Benderli2009,Strukov2008}, so that it can account for the asymmetry of the nonlinear profile characterizing the change in memristance and the varying depth of the boundary effects, as illustrated in Figure~\ref{fig:window_m_effect}.  In particular, the parameter $p$ determines the flatness of the region around $\zMax$, whereas the parameter $r$ controls the asymmetry of the window, i.e., for $0<r<2$ (excluding $r=1$) the window is left skewed, while for $r>2$ it is right skewed.  

\subsection{Relating window functions to sigmoidals}
\label{Relating sigmoidal and window functions}
We now show that the class of windows~\eqref{list:window-prop1}-\eqref{list:window-prop4} can be related to sigmoidal functions. Sigmoidal functions are elongated $S$-shaped curves, which are used as a generic representation of saturation-related phenomena in a variety of fields, from neuroscience and biology to physics. Formally, a function $S:\mathbb{R} \rightarrow \mathbb{R}$ is a \textit{sigmoidal} function if, and only if~\cite{Drakopoulos1995}:
\begin{enumerate*}
\stepcounter{listnumber}
 \renewcommand{\theenumi}{\Alph{listnumber}.\arabic{enumi}}
 \renewcommand{\labelenumi}{\Alph{listnumber}.\arabic{enumi}}
\item \label{list:sigmoidal-prop1} $S$ is monotonic; 
\item \label{list:sigmoidal-prop2} $\inf_{\mathbb{R}}S=L$, $\sup_{\mathbb{R}}S=U$. 
\end{enumerate*}
Hence as $x$ increases, the sigmoidal increases monotonically from a lower asymptotic value ($L$) towards its upper asymptote ($U$), and the slope of the curve increases until it reaches a maximum at the point of inflection ($x_\sigma$) after which it decreases. In this work, all the sigmoidals $S$ are (normalised) unit sigmoidals, i.e., bounded between $L=0$ and $U=1$. Any non-unit sigmoidal $\tilde{S}$ can be converted to its unit sigmoidal using $S(x)=(\tilde{S}(x)-L)/(U-L).$

Consider now a window function $f(z)$ complying with (\ref{list:window-prop1})-(\ref{list:window-prop4}). Then its reciprocal 
\begin{align}
\label{eq:inv_window}
\bar{f}(z) := \frac{1}{f(z)}
\end{align} 
has the following properties: 
\begin{enumerate*}
 \stepcounter{listnumber}
 \renewcommand{\theenumi}{\Alph{listnumber}.\arabic{enumi}}
 \renewcommand{\labelenumi}{\Alph{listnumber}.\arabic{enumi}}
 \item \label{list:inverse-sigmoidal-prop1} $\bar{f}(z):(0,1)\rightarrow [1/f(\zMax),+\infty)$
 \item \label{list:inverse-sigmoidal-prop2} $\lim_{z\to 0^+} \bar{f}(z)=\lim_{z\to 1^-} \bar{f}(z)=+\infty$
 \item \label{list:inverse-sigmoidal-prop3} single minimum $\bar{f}(\zMax)=1$
 \item \label{list:inverse-sigmoidal-prop4} continuous differentiability of $\bar{f}(z)$,
\end{enumerate*}
i.e., $\bar{f}(z)$ is a strictly convex function in $z\in(0,1)$ with a minimum at $z=z_\text{max}$ and asymptotes at $z=0,1$.

We can use these properties to characterise the solution of the memristive differential equation~\eqref{eqn:memristive_system_definition_special_state} governing the dynamics of the internal state variable. Using the separability of~\eqref{eqn:memristive_system_definition_special_state}, it follows that:\footnote{To remove the singularity of~\eqref{eqn:sigmoidal-in-memristor-step1} at $z=0$ and $z=1$, in practice the integral is evaluated in the range $z \in [\delta, 1-\delta]$ where $0 < \delta \ll 1$, as detailed in \cite{Shin2010a,Georgiou2013a}.}
\begin{align}
 \label{eqn:sigmoidal-in-memristor-step1}
 F(z) := \int_0^z \bar{f}(\xi) \, d\xi = \alpha \int_0^t i(\tau) \, d\tau = \alpha \left(q(t)-q_0\right),
\end{align}
where $q(t)$ is the total charge that has passed through the device taking into account the polarity of the input with $q_0:= q(0)$ the initial charge and $z=z(t)$ with $z(0)=0$. It follows from~\eqref{list:inverse-sigmoidal-prop1}-\eqref{list:inverse-sigmoidal-prop4} that $F(z)$ is a strictly increasing function since $\bar{f}(z)>0$ for $z \in (0,1)$ with a single point of inflection at $\zMax$. It is also bounded between the two asymptotes at $z=0$ and $z=1$. Furthermore, since $F(z)$ is strictly increasing, its unique inverse exists in the range of interest  $z\in (0,1)$ and is itself monotonic and bounded between $z=0$ and $z=1$. In summary, the state variable satisfies the requirements of a sigmoidal~\ref{list:sigmoidal-prop1}--\ref{list:sigmoidal-prop2}. Consequently, $F(z)$ is the inverse of a scaled and shifted sigmoidal of the charge. Let $z=S(q)$ indicate the sigmoidal curve, we can then rewrite  (\ref{eqn:sigmoidal-in-memristor-step1}) as:
\begin{equation}
 \label{eqn:sigmoidal-in-memristor-step1a}
 F(z) = \alpha \left(S^{-1}(z)-q_0\right), 
\end{equation}
and the sigmoidal of the charge is obtained 
from~\eqref{eqn:sigmoidal-in-memristor-step1a}:
\begin{equation}
\label{eqn:sigmoidal-in-memristor-step2}
 z = F^{-1}(\alpha (q-q_0) ) = S \left (\frac{F(z)}{\alpha}+ q_0 \right) = S(q). 
\end{equation}

Substituting \eqref{eqn:sigmoidal-in-memristor-step2} in \eqref{eqn:effective_resistance} and~\eqref{eqn:memristive_system_definition_special_output} gives the output voltage in terms of the input and its integral:
\begin{equation}
\label{eqn:effective_output_expression}
v= \mathcal{M}\left(S(q)\right) i(t) 
=\left[ S(q) \Ron  + \left(1-S(q) \right)\Roff \right] i(t).
\end{equation}
This expression shows that the nonlinearity of the response is encapsulated in the sigmoidal curve $S(q)$. In particular, the memristance $\mathcal{M}\left(S(q)\right)$ is a linearly transformed sigmoidal with asymptotic values corresponding to the limiting resistances of the device, $\Roff$ and $\Ron$. The sigmoidal indicates that the effect of the integrated input (i.e. charge) on the state of the system  becomes smaller closer to its two saturation boundaries ($z=0$ or $z=1$). After a limited amount of charge is injected, the system reaches saturation and behaves as a linear resistor with resistance $\Roff$ for $z=0$, or, $\Ron$ for $z=1$. Between the two saturation limits, the sigmoidal curve determines the amount of integrated input necessary to induce a change in the state of the system. Therefore, the window in the state equation enforces the boundary conditions (by limiting $z$) as well as modeling the nonlinear change of $z$.

\begin{figure*}[p]
\begin{center}
\includegraphics[width=.8\textwidth]{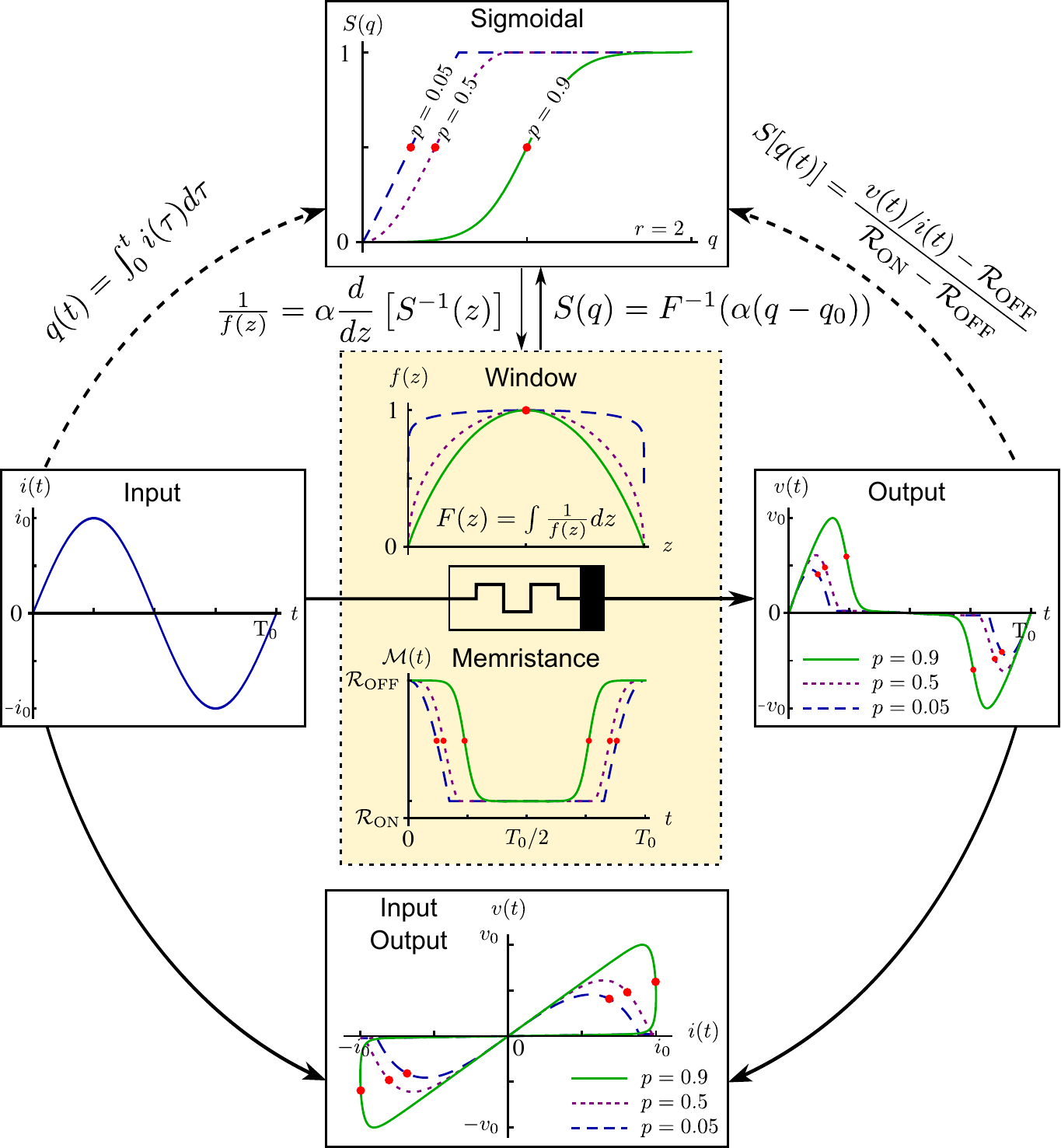}
\caption[] {
Edge effects in the response of the memristive system~\eqref{eqn:effective_resistance}-\eqref{eqn:memristive_system_definition_special} with window function~\eqref{eqn:window-new} to a sinusoidal input current, $i(t)=i_0\sin(2\pi t/T_0)$. Symmetric windows ($r=2$) with varying depths of edge effects ($p=0.05, 0.5, 0.9$) are presented together with their corresponding sigmoidal curves. The diagram illustrates the relation between the sigmoidal and the window function, as well as the input/output signals and the memristance. As seen in the sigmoidals, higher values of $p$ (which correspond to deeper edge effects) result in a higher charge threshold for the memristance. The red dots indicate the point at which the rate of change is maximal; for higher $p$, this point is delayed towards the end of the period $T_0$ and, as a result, the memristance changes noticeably later. Model and input parameters: $i_0=0.3\,\text{mA}$, $T_0=8\,\text{s}$, $\alpha=5000\,\text{C}^{-1}$, $\Ron=50\,\Omega$, $\Roff=2.5\,\text{k}\Omega$ and $z_0=10^{-4}$.
}
\label{fig:response_r_const}
\end{center}
\end{figure*}

From~\eqref{eqn:effective_output_expression} it is clear that the memristive system~\eqref{eqn:effective_resistance}--\eqref{eqn:memristive_system_definition_special} with a window~\eqref{list:window-prop1}-\eqref{list:window-prop4} is a charge-controlled, current-driven memristor as defined in~\eqref{eqn:memristor_definition}. In addition, it can be shown that, under these assumptions, the system is also an ideal memristor, as specified by~\eqref{list:ideal-memristor-properties2}-\eqref{list:ideal-memristor-properties4}.\footnote{Note that the properties (\ref{list:window-prop1})-(\ref{list:window-prop4}) of $f(z)$ assumed here are \emph{sufficient} conditions for ideal generic memristors, yet constitute a stricter subset of those proposed in \cite{Chua2015}. } To see this, integrate~\eqref{eqn:effective_output_expression} with respect to $t$ to obtain the constitutive relation in the $q-\varphi$ plane: 
\begin{equation}
\label{eqn:memristor-characteristic-equation}
\varphi(q)= \int_{q_0}^q \mathcal{M}\left(S(\xi)\right) \, d\xi.
\end{equation}
Because the memristance $\mathcal{M}\left(S(q)\right)$ is a positive function bounded between $\Ron$ and $\Roff$ the system is passive~\cite{Chua1971}. It also follows that each window satisfying (\ref{list:window-prop1})-(\ref{list:window-prop4}) corresponds to a unique $\varphi(q)$, which is a strictly increasing nonlinear function of $q$, thus satisfying all the requirements of an ideal passive memristor. We remark that in the context of Chua's new classification of memristors~\cite{Chua2014,Chua2015}, the procedure detailed above converts an ideal generic memristor into its corresponding ideal memristor.

In conclusion, although the memristive system~\eqref{eqn:memristive_system_definition_special_state} incorporates a window function to model nonlinearities and boundary effects, the separability of the state equation together with the form of the window  implies that the system is equivalent to an ideal memristor. Note that the window functions in Strukov \textit{et al}~\cite{Strukov2008}, Benderli \textit{et al}~\cite{Benderli2009}, Prodromakis \textit{et al}~\cite{Prodromakis2011} and Joglekar and Wolf~\cite{Yogesh2009}, as well as the new window proposed here in Eq.~\eqref{eqn:window-new}, fall under this class, and the corresponding memristive systems are ideal memristors. On the other hand, the window function in Ref.~\cite{Biolek2009} does not fulfil~(\ref{list:window-prop3}) and, as a result, the characteristic $q-\varphi$ curve is a double-valued function~\cite{Corinto2012} which cannot be classified as an ideal memristor. The window functions proposed by Takahashi \textit{et al}~\cite{Takahashi2014} and Corinto \textit{et al}~\cite{Corinto2012} are also excluded from our analysis since they do not comply with (\ref{list:window-prop3}). The window by Corinto \textit{et al}~\cite{Corinto2012,Ascoli2013} may also introduce step discontinuities, violating~(\ref{list:window-prop4}) and leading to a non-ideal memristor. Nevertheless, if such windows are carefully defined to avoid the introduction of discontinuities, the procedure proposed above is still applicable and can lead to an ideal memristor.

\begin{figure*}[p]
\begin{center}
\includegraphics[width=.8\textwidth]{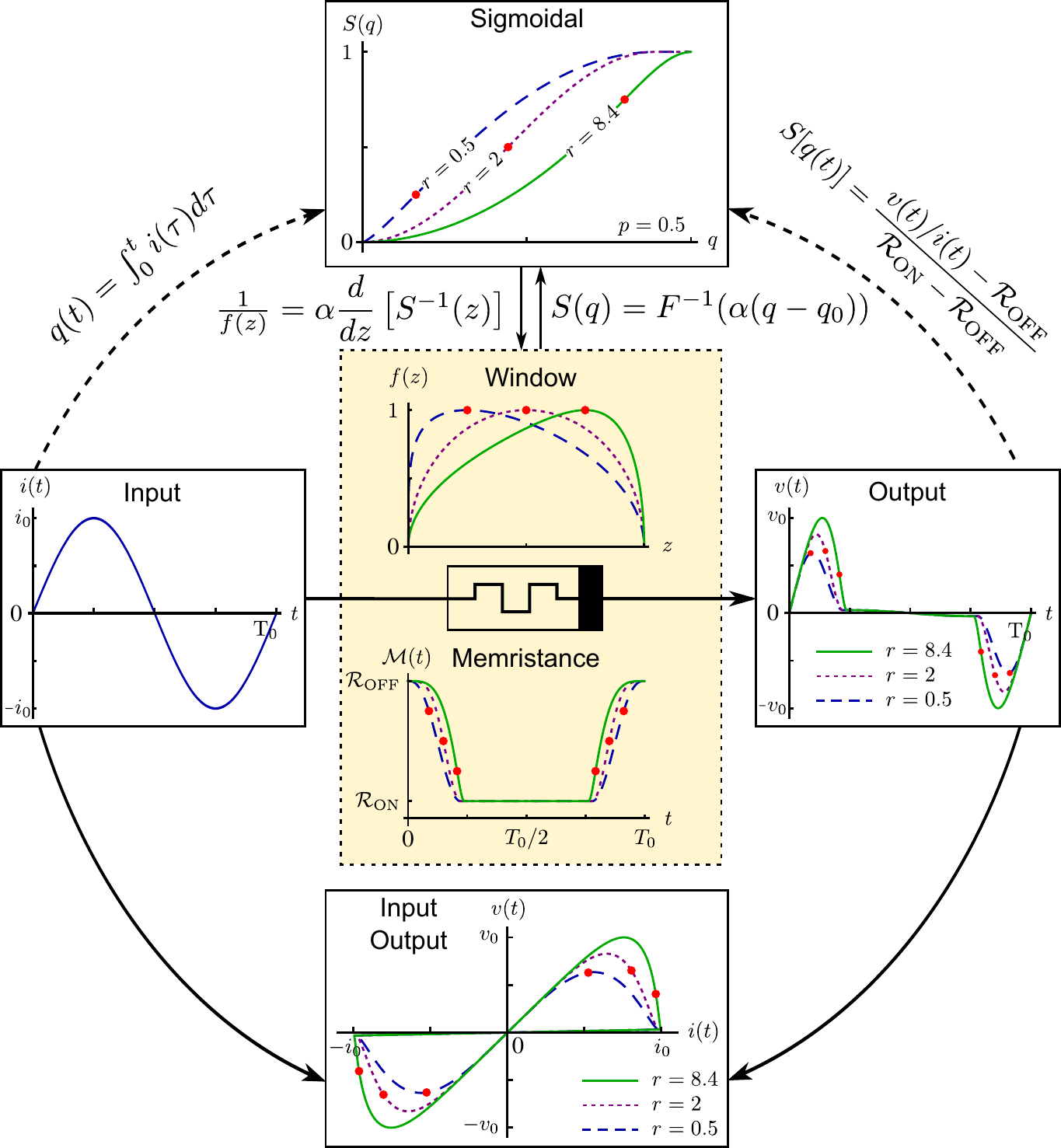}
\caption[]
{
	Asymmetry effects in the response of the memristive system \eqref{eqn:effective_resistance}-\eqref{eqn:memristive_system_definition_special} with window function~\eqref{eqn:window-new} to a sinusoidal input current, $i(t)=i_0\sin(2\pi t/T_0)$. In this case, the parameter $p$ controlling the depth of the edge effects is kept fixed at $p=0.5$ and the effect of the asymmetry is explored, with values $r=0.5$ (left skewed), $r=2$ (symmetric), and $r=8.4$ (right skewed). As in Figure~\ref{fig:response_r_const}, the diagram illustrates the relation between the sigmoidal and the window function, as well as the input/output signals and the memristance. The effect of asymmetry in the sigmoidals is to modify the position of the inflection point, which also modifies the characteristic shape of the $i-v$ curve. Model and input parameters as in Figure~\ref{fig:response_r_const}.
}
\label{fig:response_p_const}
\end{center}
\end{figure*}

\textit{\textbf{Example 1:}} We use the new generalised window~\eqref{eqn:window-new} to illustrate the connection between window functions and sigmoidals.  Following~\eqref{eqn:sigmoidal-in-memristor-step1}, we first integrate the reciprocal of~\eqref{eqn:window-new} to obtain:  
\begin{align}
\label{eq:full_F}
F(z) =&
\frac{z^{1-p\,r} \,\, {}_2{F}_1\left(p,\frac{1-p\,r}{1-r},1+\frac{1-p\,r}{1-r},z^{1-r}\right)}{(1-p\,r)  \, \left [- \psi(r) \right]^p},
\end{align} 
where ${}_2F_1 (\cdot)$ is the Gaussian hypergeometric function \cite{Abramowitz1964}. The corresponding sigmoidal can then be obtained by numerically inverting this function. In Figure~\ref{fig:response_r_const} we use these formulas to illustrate the response of the system \eqref{eqn:effective_resistance}-\eqref{eqn:memristive_system_definition_special} with window~(\ref{eqn:window-new}) to a sinusoidal input. The window is chosen to be symmetric ($r=2$) and we vary the value of the parameter $p$ determining the flatness of the window, and we also show the corresponding sigmoidals. Note how higher values of $p$, corresponding to deeper edge effects, lead to larger charge threshold for the memristance to rise. Similarly, Figure~\ref{fig:response_p_const} demonstrates how the response of the same system changes when the window (\ref{eqn:window-new}) is made asymmetric by keeping the parameter $p$ fixed while varying $r$.

If we fix $p=1$, we can obtain the sigmoidal for our window~\eqref{eqn:window-new} analytically. In this case, \eqref{eq:full_F} simplifies to
\begin{equation}
 \label{eqn:transformation-window_to_sigmoidal-example_step1}
 F(z) = \frac{\ln\left|1-z^{1-r}\right|}{(1-r)\, \psi(r)} = \alpha \, q(t),
\end{equation}
where we assume $q_0=0$ for simplicity, and the inverse of \eqref{eqn:transformation-window_to_sigmoidal-example_step1} then gives the sigmoidal:
\begin{equation}
 \label{eqn:transformation-window_to_sigmoidal-example_step2}
 S(q)=\left[1\pm e^{\alpha q (1-r)\psi(r)}\right]^\frac{1}{1-r}.
\end{equation} 

\textit{\textbf{Example 2:}} Conversely, we can convert sigmoidal curves to their corresponding window functions. This reverse transformation can be used to suggest possible window functions based on the extensive range of sigmoidals studied in the literature.

To illustrate this point, consider one of the most commonly used sigmoidals, i.e., the hyperbolic tangent standardised to the unit interval
\begin{equation}
 \label{eqn:sigmoidal_tanh_def}
 S(q)=\dfrac{1}{2}\left[\tanh(2\alpha q+\alpha_1)+1\right],
\end{equation}
where $\alpha_1\in\mathbb{R}$ is a dimensionless constant determining the horizontal shift and $\alpha \in \mathbb{R}^+ $ controls the growth rate of $S(q)$. Assuming again $q_0=0$, it follows from (\ref{eqn:sigmoidal-in-memristor-step1a}) that
\begin{equation}
 \label{eqn:transformation-sigmoidal_to_window-example_step1}
 F(z)=\alpha \, S^{-1}(z)=\dfrac{1}{2}\left(\text{artanh}(2z-1)-\alpha_1\right),
\end{equation}
which is then differentiated to obtain
\begin{equation}
 \label{eqn:transformation-sigmoidal_to_window-example_step2}
\frac{d\left[F(z)\right]}{dz}=\frac{1}{4 z (1-z)},
\end{equation}
yielding the window 
\begin{equation}
f(z)=4 z (1-z).
\end{equation} 
Hence the corresponding window for \eqref{eqn:sigmoidal_tanh_def} is a scaled version of the window in Strukov \textit{et al}~\cite{Strukov2008} and closely related to the logistic curve~\cite{Turner1976,Marusic1996}. Note that by setting $p=1$ and $r=2$ in~\eqref{eqn:window-new} it can be verified that this window is a subclass of the generalised window proposed here.

\subsection{Summary of transformations and examples}
\label{sec:Definition of transformations}

The transformations between the window function and the sigmoidal can be summarised as follows. 

Given a memristive system~\eqref{eqn:effective_resistance}--\eqref{eqn:memristive_system_definition_special} with window function $f(z)$ satisfying~\eqref{list:window-prop1}--\eqref{list:window-prop4}, then the internal state variable $z$ follows a sigmoidal curve
\begin{equation}
 \label{eqn:transformation-window_to_sigmoidal}
 z=S(q) = F^{-1}(\alpha (q-q_0)),
\end{equation}
where $F(z)=\int 1/f(z) dz$ and $z\in(0,1)$.
 
\begin{figure*}[!tb]
	\begin{center}
		\includegraphics[width=\textwidth]{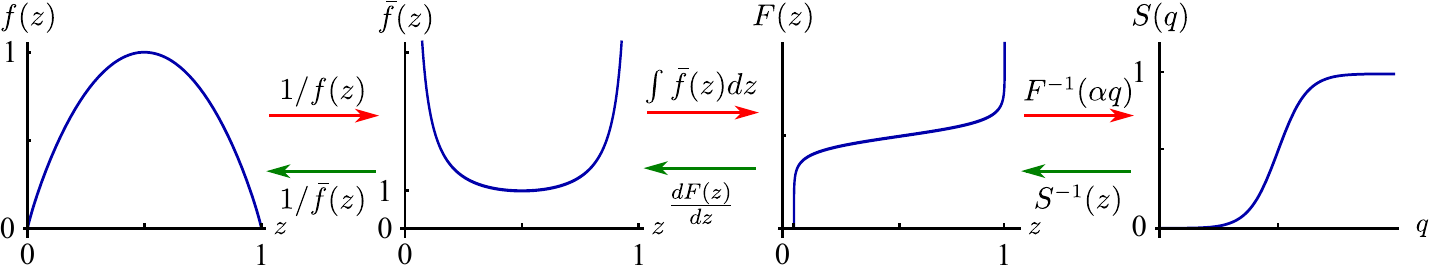}
		\caption[]{Illustration of the transformations required to convert a window function $f(z)$ to a sigmoidal $z=S(q)$ (from left to right, red arrows) and vice-versa (from right to left, green arrows).	}
		\label{fig:transformations_steps}
	\end{center}
\end{figure*}

Conversely, if the sigmoidal $z=S(q)$ is known, then we compute the corresponding window $f(z)$ by applying:
\begin{equation}
 \label{eqn:transformation-sigmoidal_to_window}
 \frac{1}{f(z)}=\alpha\,\dfrac{d}{dz}\left[S^{-1}(z) \right],
\end{equation}
which follows from inverting~\eqref{eqn:sigmoidal-in-memristor-step2} and considering~\eqref{eqn:sigmoidal-in-memristor-step1}. An illustration of these transformations is shown in Figure~\ref{fig:transformations_steps}.

These transformations allow us to obtain windows from the wide range of sigmoidal functions extensively used in the modelling of various physical and biological processes~\cite{Turner1976,Ratkowsky1986,Fekedulegn1999}. The windows obtained can then be used when fitting data, as well as giving insight to the underlying processes responsible for the observed dynamics. Without attempting to be exhaustive, we have computed windows from some of the most commonly used sigmoidal curves with different characteristics. The calculations leading to these windows follow the same pattern as above (hence not shown), and we list the results in Table~\ref{table:list-of-sigmoidals}.

\section{Window functions from experimental sigmoidals}
\label{sec:experimental procedure}

\begin{table*}[!th]
	\caption
	{Commonly used sigmoidals~\cite{Fekedulegn1999,Ratkowsky1986} and their corresponding window functions. All the sigmoidals are in unit form (see Sec.~\ref{Relating sigmoidal and window functions}) and the window functions comply with~\eqref{list:window-prop1}--\eqref{list:window-prop4}. The parameter $r$ affects various features of the sigmoidals (e.g., inflection point, growth rate, horizontal shift). The normalization $\psi(r)$ is defined such that $f(\zMax)=1$. The Weibull and Morgan-Mercer-Flodin (MMF) functions can only be viewed as sigmoidals within a restricted range of values~\cite{Fekedulegn1999,Ratkowsky1986}. Note that the charge $q$ is always scaled by an overall normalization constant $\alpha$ of appropriate dimension as in Eq.~\eqref{eqn:window-new}. 
	}
	\label{table:list-of-sigmoidals}
	\begin{tabular}{lllll}
		\hline
		\addlinespace[1mm] 
		\textbf{}  & 
		\textbf{Sigmoidal,} $S(q)$ & \textbf{Window function,} $f(z)$ & \textbf{Scaling,} $\psi(r)$ & \textbf{Parameters} \\
		\addlinespace[1mm] 
		\hline
		\addlinespace[2mm]
		
		\rotatebox[origin=c]{90}{Eq.~\eqref{eqn:window-new}} &
		Inverse of~Eq.~\eqref{eq:full_F} & $  \left[\psi(r) \, z (1-z^{r-1})\right ]^p$ & $\dfrac{r^{\frac{r}{r-1}}}{r-1}$& \parbox{.15\textwidth}{ \raggedright  $r \in {\mathbb{R}}^+ -\lbrace 1 \rbrace$, \\ $p\in(0,1]$} \\ 
		\addlinespace[3mm]
		
		\rotatebox[origin=c]{90}{Eq.~\eqref{eqn:window-new}} &
		$\left[1\pm e^{(1-r) \psi(r) \, \alpha q} \right]^\frac{1}{1-r}$  & $\psi(r) \, z (1-z^{r-1})$ & $\dfrac{r^{\frac{r}{r-1}}}{r-1}$& \parbox{.15\textwidth}{\raggedright $r \in {\mathbb{R}}^+ -\lbrace 1 \rbrace$, \\ $p=1$} \\ 
		\addlinespace[3mm]
		
		\rotatebox[origin=c]{90}{Gompertz} &
		$ e^{-r e^{\left(-e \, \alpha q\right)}}$ & $-e \, z \ln\left(z\right)$ & {\centering ---} & $r>0$ \\ 
		\addlinespace[3mm] 
		
		\rotatebox[origin=c]{90}{Arctan} &
		$\dfrac{1}{2} + \dfrac{1}{\pi} \arctan\left(\pi \, \alpha q\right)$ & $\sin^2(\pi z)$ & {\centering ---} & {\centering ---} \\ 
		\addlinespace[3mm] 
		
		\rotatebox[origin=c]{90}{Weibull} &
		$1-e^{\left[-\psi(r) (\alpha q)^r\right]}$ & $\dfrac{r \psi(r)}  {(1-z)^{-1}} \left[-\dfrac{\ln(1-z)}{\psi(r)}\right]^{1-\frac{1}{r}}$ & $\dfrac{e^{r-1}}{r \, (r-1)^{r-1}}$ & \parbox{.15\textwidth}{ \raggedright $-1\le r <0$,  $r>1$} \\ 
		\addlinespace[3mm]
		
		\rotatebox[origin=c]{90}{MMF} &
		$\dfrac{1}{1+ \psi(r) (\alpha q)^{-r}}$ & $r \left[\psi(r) \dfrac{z^{1-r}}{(1-z)^{r+1}} \right]^{-1/r}$ & $\dfrac{(r+1)^{r+1}}{(r-1)^{1-r} (4 r )^{r}}$ & $r>1$ \\
		\addlinespace[3mm]
		
		\hline
	\end{tabular}
\end{table*}

The foundations laid above suggest a series of steps through which suitable window functions can be fitted to experimental data under the assumptions that the measured device will conform to the class of windowed memristors studied above. In particular, the experimental input-output data can be used to extract the sigmoidal of the time-integral of the input $z=S(q)$, which can then be related to the window function by using the transformation~\eqref{eqn:transformation-sigmoidal_to_window}. This reconstructed window function empirically models the nonlinearity of $\dot{z}$, and can be tailored to the specific device under investigation. In particular, different sigmoidals could be fitted and compared to select the most appropriate description and window function. Conversely, if a physical model exists to inform the choice of window, the transformation~\eqref{eqn:transformation-window_to_sigmoidal} can be used to obtain the corresponding sigmoidal, which can then be fitted to the data. 

The relation between the sigmoidal, its corresponding window function, and the underlying input-output behaviour of the device is illustrated in Figures~\ref{fig:response_r_const} and~\ref{fig:response_p_const}. The internal state $z$ is an abstract variable which in general is not directly related to any measurable quantities unless certain aspects of the system are pre-specified. However, if we assume that the memristance is modelled by \eqref{eqn:effective_resistance} then $z$ can be mapped to measurable quantities. As described above, the cumulative effect of the input becomes smaller the closer the system is to any of the boundaries ($z=0$ or $z=1$) at which the system reaches saturation and behaves as a linear resistor with resistance $\Roff$ or $\Ron$, respectively. Hence, while in saturation, the integrated input has no effect on the resistance of the device. This leads us to consider a normalised instantaneous time-dependent resistance:
\begin{equation}
\label{eqn:z-is-macroscopic}
z(t)= \frac{\left[v(t)/i(t)\right]-\Roff}{\Ron-\Roff},
\end{equation}
which scales the resistance with respect to the minimum ($\Ron$) and maximum ($\Roff$) resistance states of the device. These limiting states need to be determined by driving the system to both saturation limits in a sustained manner with two input signals of opposite polarity. This process calibrates the range $z\in[0,1]$ by mapping the range of memristances to $\mathcal{M}\in[\Roff,\Ron]$. 

Once $\Ron$ and $\Roff$ have been obtained, we initialise the device to $\Ron$ or $\Roff$ and apply an input $i(t)$ measuring the instantaneous output voltage $v(t)$ at sampling times $t_i, \, i=1,\ldots,N$. Any type of input can be used provided it forces the device to explore the entire range of the memristance. For each sampling time, we evaluate the accumulated charge $q(t) = \int_0^{t_i} i(\tau)\, d\tau$ and the state variable $z(t_i)$ using \eqref{eqn:z-is-macroscopic}. We then fit the $N$ sampled pairs $(q(t_i),z(t_i))$ to a suitable sigmoidal function, such as those in Table~\ref{table:list-of-sigmoidals}. The data-fitted sigmoidal can then be converted to its corresponding window function by applying the transformation \eqref{eqn:transformation-sigmoidal_to_window}.

\section{Discussion}
\label{sec:discussion}

Window functions are commonly used in the macroscopic modelling of memristive devices in order to incorporate edge effects and other nonlinearities in a semi-empirical manner. Here we have shown that, under quite general assumptions for the window function and the memristive system, such devices are characterised by an internal state variable that exhibits a sigmoidal response to the time integral of the input. The characteristic sigmoidal of the device is linked to the underlying window function via a bidirectional transformation that can be used to obtain the corresponding sigmoidal for a given window and vice versa. 

The sigmoidal provides a description which is more easily linked to measured observables and which takes into account the effect of the input signal on the state of the system. Specifically, the sigmoidal curve can be extracted from input/output measurements and reflects the accumulated effect of the input and the saturation of the response. Such sigmoidal functions fitted to experimental data can be converted to a window function to obtain a memristive model tailored to the experimental device under investigation. Within our framework, a variety of plausible window functions can be obtained using the wide range of sigmoidal functions for their potential use in mesoscopic models. Conversely, based on the physical properties of the device, a window function maybe favoured. In that case,  the corresponding sigmoidal can be computed and fitted to the data.

In addition, we have proposed a generalised window which extends some of the features of previously proposed windows~\cite{Yogesh2009,Benderli2009,Prodromakis2011} to allow for additional flexibility when fitting data. The proposed window is related to a known family of functions, the \textit{generic growth functions}, from which most of the commonly used sigmoidals can be derived, including the logistic and Richards curves~\cite{Turner1976,Marusic1996}.

Finally, we have shown that, under the general assumptions considered here, the incorporation of a \textit{separable} window function for modelling nonlinearities and boundary effects results in a system which is still an \textit{ideal} memristor. In contrast with Ref~\cite{Biolek2014}, where ideality is assessed based on whether a memristive system can be converted to the canonical form on the $i-v$ plane, here we assess ideality based on the properties \eqref{A.1}--\eqref{A.3} of the $q-\varphi$ characteristic curve. However, practical devices are far from ideal and some of these assumptions will not hold. For instance, the window introduced by Biolek \textit{et al}~\cite{Biolek2009} does not comply with \eqref{list:window-prop1}--\eqref{list:window-prop4} because the change in the memristance is assumed to depend on the polarity of the input. More complex dependencies have also been reported elsewhere~\cite{Strukov2009a,Shin2010a,Eshraghian2012,Kvatinsky2013}. Further work will explore the application of the procedure sketched in Section~\ref{sec:experimental procedure} to data from real memristive devices. Since the method is expected to reveal the underlying sigmoidal of each device, non-sigmoidal behaviour in real data will indicate the extent to which non-idealities play a role in different devices, and how to extend our proposed methodology of analysis. A potential extension of our methodology to non-ideal devices is to treat them as `piecewise ideal', by splitting their $q-\varphi$ curve into piecewise domains over which they are ideal. Such analysis may reveal more complex mechanisms of saturation beyond those modelled using window functions compliant with (\ref{list:window-prop1})-(\ref{list:window-prop4}), so that an upper and lower limit to the memristance is present~\cite{Georgiou2013a,Georgiou2014}. An investigation of the existence of underlying sigmoidals in \textit{experimental} memristive devices, where these ideas of piecewise ideality are explored,  will be reported elsewhere.

\section*{Acknowledgements}

This work was supported by the Engineering and Physical Sciences Research Council (EPSRC) [grant number EP/L504786/1].


\bibliography{Bibliography}

\end{document}